\documentclass[draftcls, 12pt,onecolumn,oneside]{IEEEtran}

\usepackage{}
\usepackage{amsmath}
\usepackage{amsfonts}
\usepackage{mathrsfs}
\usepackage{amssymb}
\usepackage{bm}
\usepackage{cases}
\usepackage{stmaryrd}
\usepackage{mathrsfs}
\usepackage{bbm}
\usepackage[dvips]{graphicx}
\usepackage{CJK}
\usepackage{amsmath}
\usepackage{flushend}
\usepackage{algorithm}
\usepackage{algorithmic}
\usepackage{multirow}
\usepackage{array}
\usepackage{graphicx}
\usepackage{setspace}
\usepackage{subfigure}
\usepackage{epstopdf}
\usepackage{float}
\usepackage{cite}
\usepackage{enumerate}
\usepackage{xcolor}

\begin{document}

\title{\begin{center}UAV-Involved Wireless Physical-Layer Secure Communications: Overview and Research Directions\end{center}
\author{
	Hui-Ming~Wang,~\IEEEmembership{Senior Member,~IEEE,}~%
	Xu~Zhang,
	and~Jia-Cheng~Jiang%
}
\thanks{Hui-Ming Wang, Xu Zhang and Jia-Cheng Jiang are with the School of Electronics and Information Engineering, and also with the MOE Key Laboratory for Intelligent Networks and Network Security, Xi'an Jiaotong University, Xi'an, 710049, China. Email: {\tt xjbswhm@gmail.com, jcx8008208820@stu.xjtu.edu.cn, j1143484496b@stu.xjtu.edu.cn}.
}}
\maketitle

\begin{abstract}
Due to their flexible deployment and on-demand mobility, small-scale unmanned aerial vehicles (UAVs) are anticipated to be involved in widespread communication applications in the forthcoming fifth-generation (5G) networks. However, the confidentiality of UAV communication applications is vulnerable to security threats due to the broadcast nature and dominant line-of-sight (LoS) channel conditions, and physical-layer security (PLS) technique can be applied for secrecy performance enhancement in such a context. On the other hand, it is also promising to exploit UAVs to cooperatively protect secure communications. This article provides an overview of the recent research efforts on UAV-involved secure communications at the physical layer. We focus on the design of secure transmission schemes according to different roles of UAVs and the optimization of introduced degrees of freedom (DoFs) by the unique characteristics of UAVs. We also propose some future research directions on this topic.

\end{abstract}


\IEEEpeerreviewmaketitle

\section{Introduction}
Equipped with various kinds of sensors and actuators like Inertial Measurement Unit (IMU), range sensors (ultrasonic, infrared, laser), barometer, magnetometer, Global Position System (GPS), cameras and visual systems, unmanned aerial vehicles (UAVs) are promising to support a wide range of applications due to their characteristics of flexible deployment, low acquisition and maintenance costs, high maneuverability and hovering ability \cite{UAVTutorial1}. Historically, UAVs have been considered for military applications from the beginning to carry out some simple but risky tasks, such as monitoring and attacking the hostile targets. Whereafter, further attentions have been paid to applying small-scale UAVs for emerging civilian tasks, including aerial photography, emergency search and rescue, resource exploration and cargo transport, etc. The U.S. Federal Aviation Administration (FAA) has released the operational rules to guideline the working definition of low-altitude small-scale UAVs with aircraft weight less than 55 pounds and maximum altitude less than 400 feet above ground level. Whereafter, FAA has launched a further national program, namely ``Drone Integration Pilot Program'', to explore the expanded use of UAVs. Motivated by their unique characteristics, UAVs have also been considered to play an important role in the future communication systems \cite{UAVWirelessCommunSurvey1}. On one hand, since UAVs are generally exploited to carry out tasks at a relatively high altitude, the aerial-to-ground (A2G) line-of-sight (LoS) channels are likely to provide channel superiority compared with ground communication channels in cellular networks, which are significantly affected by severe fading and shadowing effects. On the other hand, due to their on-demand mobility characteristic, UAVs can be flexibly deployed which will introduce the new degrees of freedom (DoFs) with respect to their positions. To facilitate efficient and reliable transmission, the established UAV-involved communication links can be categorized as payload communication and control and non-payload communication (CNPC), the specific requirements of which can be totally different and thus have been specified by the 3rd Generation Partnership Project (3GPP) recently.

Due to the openness of wireless environment, the security and privacy of wireless communication applications are of utmost concern. Particularly, the confidentiality of UAV wireless communications is more challenging to be protected under LoS propagations, which potentially provide strong quality of A2G wiretap channels for the hostile. Therefore, effective methods are urgently needed for secure UAV communications. Traditionally, the cryptography-based methods are exploited to protect the confidentiality of secure communications by using shared secret keys. However, the high mobility characteristic of UAVs makes the corresponding key management and distribution more challenging. Besides, ultra-reliable and low-latency communication links between UAVs and the associated ground control stations (GCSs) are required to support their two-way on-demand control to ensure safe and efficient operation of UAVs. Therefore, the cryptography-based methods are unsuitable due to the significant processing delay. In addition, a general drawback of cryptography-based methods is that these methods are dependent on the computational complexity, and thus the perfect secrecy cannot be guaranteed. The methods will be invalid if the hostile has powerful computing devices. Under this condition, physical-layer security (PLS) has been proposed and developed as a key complementary technique for secure wireless communications \cite{PLSWork}. The basic idea of PLS is to exploit the randomness characteristics of wireless channels, which is key-less and thus promising for UAV secure communications to overcome the aforementioned drawbacks. In addition, since secrecy performance is highly dependent on the superiority of legitimate channels to wiretap channels, additional DoFs provided by the on-demand mobility of UAVs can be exploited to guarantee the expected channel superiority and thus improve secrecy performance.

In such a context, UAV-involved physical-layer secure wireless communications has attracted increasing interest in recent years. In general, they could be categorized as UAV-enabled secure communications and UAV-aided secure cooperation, according to different roles of UAVs in secure communications. Specifically, UAVs can be exploited as legitimate transceivers for the former typical scenario to establish direct communication links, such as aerial base stations providing temporary wireless connections for ground users or aerial terminals carrying on other specific tasks, e.g. surveillance. While for the latter typical scenario, UAVs are exploited to enable friendly relaying or jamming, so as to cooperatively enhance secrecy performance of secure communications. Compared with the existing traditional PLS researches, the most significant difference is that, extra security DoFs have been introduced in UAV wireless communications, inspired by the flexible deployment and on-demand mobility of UAVs. Especially, three-dimensional (3D) deployment of static UAVs, and trajectory design of mobile UAVs, together with transmit power can be jointly optimized for secrecy performance enhancement.
Efficient secure transmission schemes should be carefully designed and position/trajectory/power should be optimized.
However, a well-organized overview of recent research progress on this topic is still absent as far as we know, which motivates this article.

The article is organized as follows. The typical application scenarios of UAV-involved secure wireless communications as well as the main design considerations brought in by the mobility of UAVs
are introduced in Section \uppercase\expandafter{\romannumeral2}. Then, we focus on the recent research efforts on UAV-involved secure communications under various typical application scenarios in Section \uppercase\expandafter{\romannumeral3}-Section \uppercase\expandafter{\romannumeral5}. Finally in Section \uppercase\expandafter{\romannumeral6}, we propose some future research directions on this topic.

\section{Typical Application Scenarios and Main Design Considerations}
\begin{figure}[t]
\centering
\includegraphics[width=12cm,height=6cm]{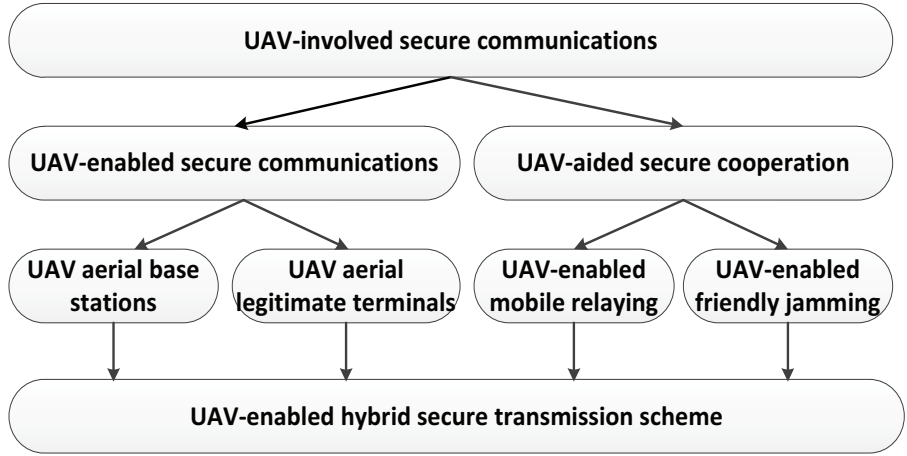}
\caption{Specific applications for UAV-involved secure communications.}\label{P1}
\end{figure}

As aforementioned above, the typical application scenarios can be essentially summarized as \emph{UAV-enabled secure communications} and \emph{UAV-aided secure cooperation} according to different roles of UAVs in secure communications. Furthermore, \emph{hybrid secure transmission scheme} could be obtained by combining the two typical application scenarios. Fig. \ref{P1} depicts the  methodology of UAV-involved secure communications.

For the first typical scenario, UAVs can be exploited as \emph{aerial base stations} to potentially provide temporary connectivity services for the area without cellular infrastructure coverage. It may happen due to natural disasters in emergency situations, or data traffic offloading in the hotspot area with densely-distributed users, etc. On the other hand, UAVs can also be exploited as \emph{aerial terminals} with their own missions such as aerial surveillance and video streaming. UAV aerial terminals are under the control of associated GCSs through CNPC links for automatic operation and are willing to establish data transmission links with ground base stations (GBSs) for information exchange. In both cases, however, the confidentiality of  UAV communications is vulnerable to security threats of the hostile due to the broadcast nature of wireless transmissions. As a result, enhancing the corresponding secrecy at the physical layer becomes critical for the above application scenarios.

To further improve secrecy performance, UAV-aided secure cooperation is recognized as another typical application scenario and can be generally categorized as \emph{UAV-enabled mobile relaying} and \emph{UAV-enabled friendly jamming}. Due to their on-demand mobility, UAVs could be deployed as mobile relays to enhance the superiority of legitimate channels in the dynamic environment, and the dominant LoS conditions can also be exploited for UAV-enabled friendly jamming to effectively degrade the quality of wiretap channels. As a result, both will improve secrecy capacity according to the basic principles of PLS.

The most significant observation is that, new security DoFs, including \emph{3D position design of static UAVs} and \emph{trajectory design of mobile UAVs}, have been introduced by the on-demand mobility characteristic of UAVs. Together with the traditional DoF, \emph{transmit power design}, they could be exploited jointly to enhance secrecy performance of UAV-involved transmission schemes under different typical application scenarios. The corresponding issues on these aspects are discussed in the following parts, respectively.

Under the secure communication scenarios involving static UAVs, the 3D position design will significantly affect secrecy performance of secure transmissions. For a tractable analysis, the optimal 3D position can be determined by jointly design of UAV horizontal position and flight altitude. The design of horizontal position is related to the distributions of both legitimate users and ground eavesdroppers. Intuitively, UAVs should be positioned horizontally close to legitimate users and far away from potential eavesdroppers when providing communication services or cooperative relaying, while in a reverse manner when providing friendly jamming. On the other hand, with the increase of UAV flight height, the effect of large-scale path loss is enhanced while the probability of the LoS path being blocked is reduced. Therefore, the optimal flight altitude generally exists under different conditions.

Due to their on-demand mobility characteristic, the trajectory design of mobile UAVs is further considered to be exploited for secure communications. It is expected to enhance the superiority of legitimate channels to wiretap channels by proper trajectory design, which is beneficial to improve secrecy capacity according to the basic principles of PLS. To facilitate the trajectory design, the periodical flight duration is generally discretized into multiple short time slots and mobile UAVs are approximately assumed static in each time slot. It is worth noting that the length of each time slot should be carefully chosen since a short time slot will simultaneously lead to high approximation accuracy of static UAVs and high complexity of the designing problems.

Combining with the positions and trajectories of UAVs, the power-domain DoFs should be jointly optimized to further improve secrecy performance. The quality of both legitimate channels and wiretap channels is time-variant during the flight period, which significantly affects the transmit power design. Generally under different application scenarios of UAV-involved secure communications, the purpose of transmit power design is to sufficiently make use of the channel superiority of legitimate channels to improve the secrecy capacity or to increase the channel superiority in an opportunistic manner.

Based on the aforementioned application scenarios and main design considerations, the recent research efforts on both UAV-enabled secure communications and UAV-aided secure cooperation are provided subsequently. After that, we will provide some valuable future research directions according to the comprehensive analysis of the recent research efforts.

\section{UAV-Enabled Secure Communications}
In this section, we provide the recent research efforts on UAV-enabled secure communications. Based on different roles of UAVs, the specific applications in this scenario can be separately categorized as \emph{UAV aerial base stations} and \emph{UAV aerial legitimate terminals}, which will be discussed in the following parts.

\begin{figure}[t]
\centering
\includegraphics[width=12cm,height=6cm]{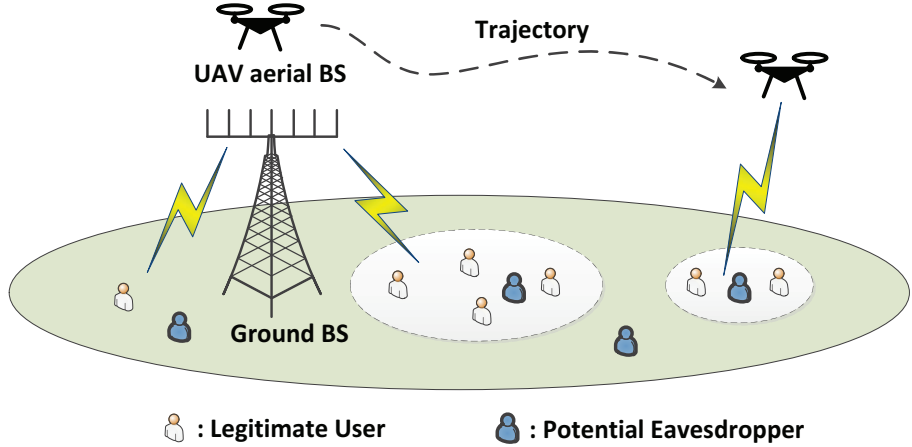}
\caption{UAV aerial base stations to provide temporary secure communication services for the certain areas.}\label{P2}
\end{figure}

\subsection{UAV Aerial Base Stations}
Due to their high mobility, UAVs can be flexibly deployed as aerial base stations to provide temporary communication services for a certain area, as depicted in Fig. \ref{P2}. Due to the broadcast nature of wireless transmissions and the dominating strong LoS channels, there exist severe eavesdropping threats on secure transmissions. Based on the location information of legitimate users and potential eavesdroppers, the on-demand mobility of UAVs can be exploited to simultaneously enhance the quality of legitimate channels and reduce the quality of wiretap channels as much as possible, and thus improve secrecy performance.

\begin{figure}[t]
\centering
\includegraphics[width=15cm,height=9cm]{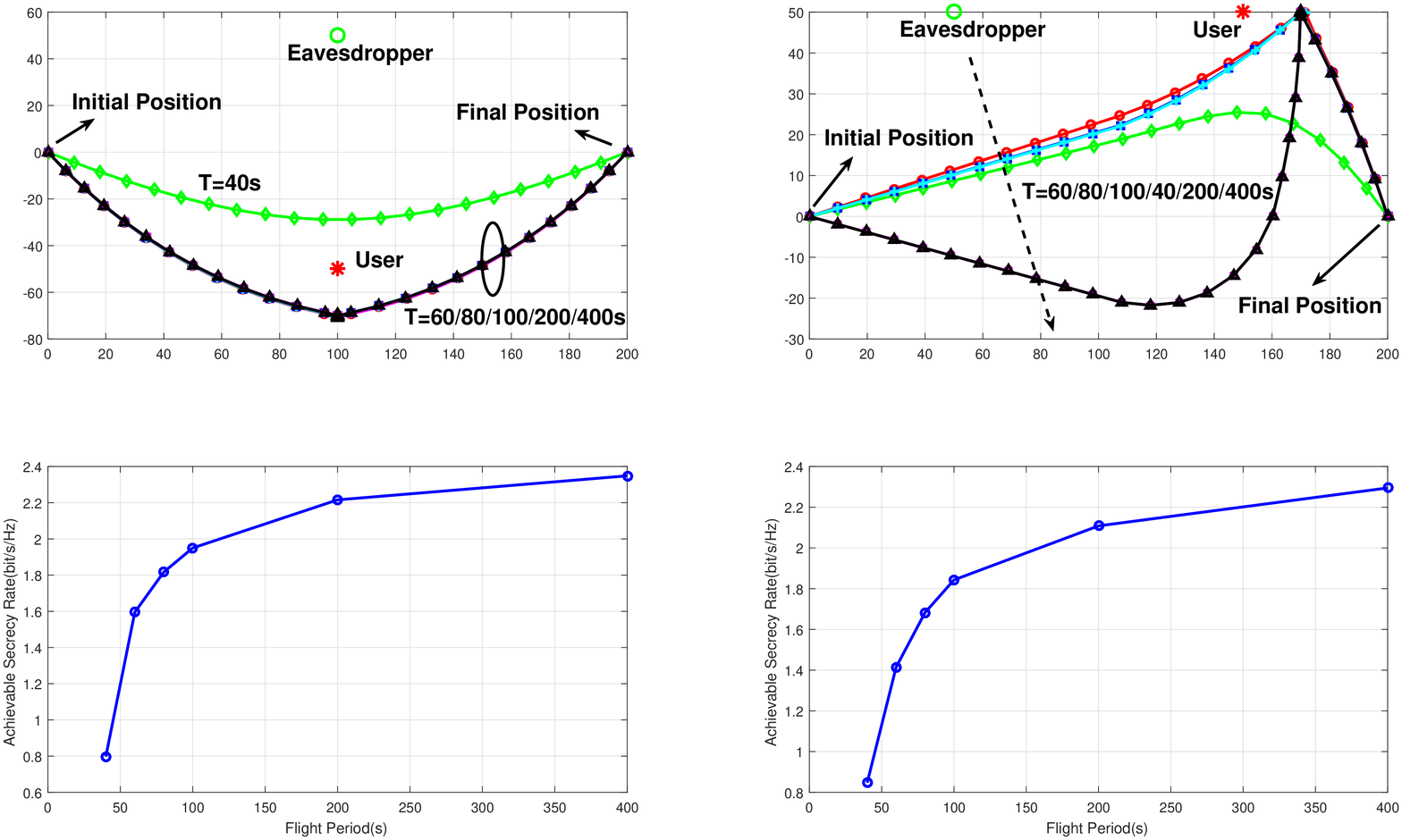}
\caption{The optimal trajectory designs of UAV aerial base station and the corresponding achievable secrecy rates against external eavesdropper for different flight periods. The maximum speed of UAV is 5m/s and the reference signal-to-noise ratio (SNR) at the reference distance ${{d}_0} = 1m$ is 80dB. The flying altitude of UAV is set as 50m.}\label{P11}
\end{figure}

The corresponding problem involving UAV aerial base stations was firstly investigated in \cite{UAVBaseStation1}. Assuming that the position of the ground eavesdropper was available under LoS propagations, the trajectory and transmit power of the UAV aerial base station were jointly optimized to maximize the average achievable secrecy rate. Fig. \ref{P11} depicts the optimal UAV trajectory designs and the corresponding achievable secrecy rates for different flight periods. It is observed that the flight period is of great importance on the feasibility in designing an efficient trajectory. If the flight period is sufficiently large, the optimal fly-hover-fly scheme is adopted where UAV aerial base station always flies at the maximum speed to reach the optimal position and then hover (for rotary-wing UAV without minimum flying speed limit) for better secrecy performance.

However, the actual locations of passive eavesdroppers are difficult to be accurately estimated in practice. Therefore, how to deal with the position uncertainty becomes more challenging. In \cite{UAVBaseStation3}, only the estimated locations of the ground eavesdroppers are available and their exact locations were assumed in an area with a bounded estimation error. In this case, the authors jointly optimized the trajectory and transmit power to maximize the average worst-case secrecy rate. To effectively solve the non-convex optimization problem, $S$-procedure was introduced to deal with the channel uncertainty problem, and the block coordinate descent method with successive convex approximation (SCA) was exploited to iteratively obtain a sub-optimal solution. The proposed robust transmission scheme is validated to significantly improve the secrecy performance in the case of imperfect location estimation.


As for static UAV aerial base stations, the multi-antenna technology is potential to enhance the superiority of equivalent legitimate channels by proper beamforming design. In \cite{UAVBaseStation2}, the authors investigated millimeter wave (mmWave) secure transmissions in Nakagami-$m$ fading environment with mixed LoS/non-line-of-sight (NLoS) A2G channels, and the 3D antenna gains of UAV aerial base stations were considered in a stochastic geometry framework. To incorporate the UAV minimum separation distance requirements, the mat\'{e}rn hardcore process was used to characterize the locations of UAV aerial base stations. In addition, part of random distributed UAVs were further exploited as friendly jammer to improve the secrecy performance, and the analytical expressions of the target user's average secrecy rate were derived. It has been shown that, though the achievable rates of both legitimate users and eavesdroppers increase with the transmit power of UAV aerial base stations, there exists an optimal transmit power for maximizing the average secrecy rate under certain conditions.


\subsection{UAV Aerial Legitimate Terminals}
Due to their flexible deployment and high mobility, UAVs have also been exploited as aerial terminals to carry out some special tasks.
Since UAV aerial terminals are practically operating in an automatic manner under the control of associated GCSs via wireless links, their communication security is a critical issue and needs protection.


In \cite{UAVLegitimateNode1}, the authors discussed the ground-to-aerial (G2A) secure communications of the static UAV legitimate terminal in the presence of a full-duplex ground eavesdropper, where the eavesdropper simultaneously performed eavesdropping and malicious jamming. For secrecy performance enhancement, the null-space based artificial noise was exploited to reduce the quality of wiretap channels. Under the condition that only the statistical channel state information (CSI) of the eavesdropper was known, the hybrid outage probability combining both transmission outage probability and secrecy outage probability (SOP) was derived. Based on the analytical expressions, the optimal power allocation policy was obtained by a bisection search. The secrecy performance could be improved by increasing the transmit power and/or equipping more transmit antennas at the source. In addition, there exists an optimal operation height of the UAV aerial legitimate terminal under different conditions according to simulation results.

To further take the mobility of UAVs into consideration,
the authors in \cite{UAVLegitimateNode2} investigated the trajectory planning of the UAV aerial legitimate terminal against malicious jamming to enhance the quality of G2A secure communications. The positions and jamming powers of ground attackers were assumed to be fixed during the flight period, and were empirically estimated by the corresponding statistical values. Then, the 3D trajectory was optimized to maximize the achievable throughput over the flight period, and SCA methods were exploited to overcome the non-convexity of the optimization problem. Moreover, the closed-form solution of the optimized 3D deployment of the static UAV aerial legitimate terminal was geometrically derived, which was an asymptotic case of the trajectory design with unlimited UAV speed. It is worth noting that the optimal hovering position for the trajectory design is relatively close to the optimal deployment of the UAV aerial legitimate terminal.

\section{UAV-Aided Secure Cooperation}
The recent research efforts on UAV-aided secure cooperation are provided in this section. According to different roles of UAVs, the specific applications in this scenario can be separately categorized as \emph{UAV-enabled mobile relaying} and \emph{UAV-enabled friendly jamming}, which will be discussed in the following parts.

\subsection{UAV-Enabled Mobile Relaying}

\begin{figure}[t]
\centering
\includegraphics[width=12cm,height=6cm]{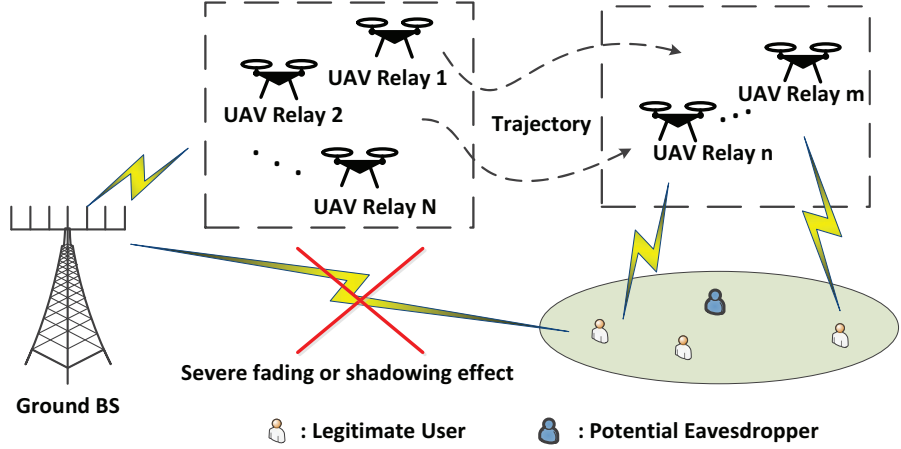}
\caption{UAV-enabled mobile relaying under the scenario where the ground direct transmission link is severely blocked by fading and shadowing effect.}\label{P3}
\end{figure}

According to the basic principles of PLS, cooperative relaying can be exploited to achieve the superiority of legitimate channels to wiretap channels, which is critical to improve secrecy capacity. However, the locations of traditional ground relay nodes are time-invariant, which makes it difficult for them to flexibly adapt to the dynamic environment. On the contrary, UAVs can move close to the related nodes for secure communications due to their on-demand mobility when serving as relay nodes, as depicted in Fig. \ref{P3}. In such scenarios, the quality of legitimate A2G channels can be enhanced while that of wiretap channels may be degraded due to their distinct geometrical locations. As a result, UAV-enabled mobile relaying is promising to further improve secrecy performance.

UAV-enabled mobile relaying was firstly investigated in \cite{UAVRelayPLS1} to maximize the average achievable secrecy rate by optimizing the transmit power allocation among the flight period. Then, a similar idea was extended in \cite{UAVRelayPLS2} to further include the joint design of dynamic UAV trajectory. It is worth noting that decode-and-forward (DF) mode is adopted in the above works, where UAV mobile relays are promising to approach GBSs for better decoding capability. However, due to their limited velocity, the adopted DF-mode UAV relays are implicitly required equipping with a large buffer to store the decoded messages, which causes significant transmission delay. Instead, amplify-and-forward (AF) mode is an effective choice for UAV-enabled mobile relaying with less processing complexity and transmission delay. Moreover, AF mode is appropriate for untrusted relay scenarios to avoid additional secrecy issues. However, the mobility characteristics of UAVs cannot be fully utilized since retransmission occurs at the adjacent time slot, and the amplified background noise will lead to relative performance loss at high SNR regime compared with DF mode. Above all, the best relaying mode can be different and is highly dependent on the specific application scenario.

Under the scenario where there exist multiple mobile UAV relays, the authors in \cite{UAVRelayPLS4} investigated the opportunistic relaying in the presence of multiple UAV eavesdroppers. Specifically, the optimal UAV relay was chosen according to the principle of maximizing the end-to-end SNR, which is observed as a random selection from the perspective of eavesdroppers. Considering the wireless backhaul reliability  from the GCS to the UAV-transmitter under Nakagami-$m$ fading conditions, the closed-form expression of SOP was then derived under the assumption that the maximum ratio combining (MRC) was applied among multiple UAV eavesdroppers. However,  the significant mobility characteristic of UAV-enabled mobile relaying was not exploited in this work. Except for relay selection, cooperative beamforming is another choice to enhance the quality of equivalent legitimate channels to improve secrecy performance. To our best knowledge, related work on this topic is still missing, which is an open issue for future researches.

\subsection{UAV-Enabled Friendly Jamming}

\begin{figure}[t]
\centering
\includegraphics[width=12cm,height=6cm]{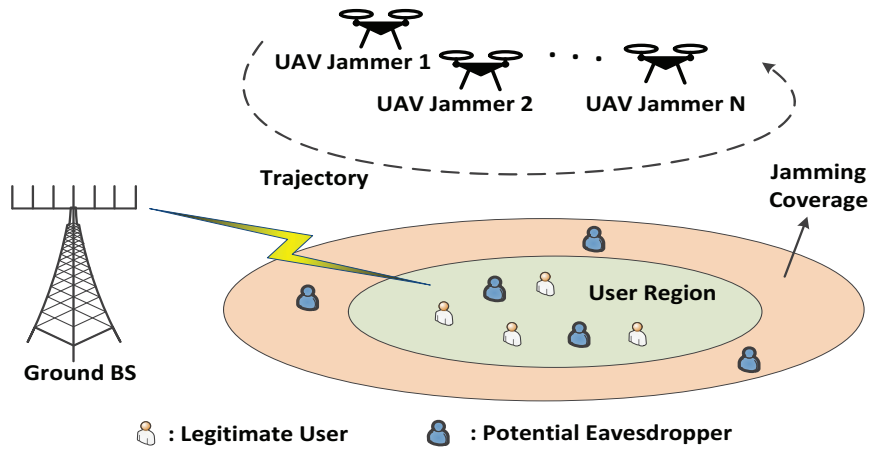}
\caption{UAV-enabled friendly jamming to transmit artificial noise against potential eavesdroppers in the certain area.}\label{P4}
\end{figure}

Except for relaying the expected messages, UAV-enabled friendly jamming is another way to cooperatively improve the security by transmitting artificial noise, as depicted in Fig. \ref{P4}. Compared with traditional ground jamming, UAVs are able to adjust their jamming power dynamically according to their relative positions with legitimate users and potential eavesdroppers due to their flexible deployment and high mobility. In addition, the dominant strong LoS A2G channels are more beneficial for jamming. On one hand, the quality of wiretap channels is significantly degraded since LoS A2G channels are less impaired by fading and shadowing effects compared with ground jamming channels. On the other hand, the CSI of A2G jamming channels is easier to obtain under LoS conditions since it is highly dependent on the relative distances between UAV mobile jammers and ground eavesdroppers. Therefore, UAV-enabled friendly jamming is potential for secrecy performance enhancement.

The basic idea of UAV-enabled friendly jamming was discussed in \cite{UAVJammingPLS1}. The average achievable secrecy rate was maximized by jointly designing the trajectory and transmit power of an UAV jammer. It is observed that the time-variant positions are determined by the flight period of UAV mobile jammer, which has an important impact on the optimal design of UAV trajectory and thus significantly affects secrecy performance. In addition, it is worth noting that an improper jamming power design is even harmful to secrecy performance since the quality of legitimate channels can be more severely degraded.

In practice, the exact locations of eavesdroppers are generally difficult to be perfectly estimated. The estimation error has a significant impact on the design of secure transmission schemes. Taking this into consideration, a potential eavesdroppers' target area model was established in \cite{UAVJammingPLS5} for UAV-enabled friendly jamming scheme design.
Based on different assumptions on A2G channels, the mixed LoS/NLoS model was adopted to characterize the average path loss of UAV jamming links. Then, the jamming coverage was determined by comparing the location-dependent interception probability of the potential eavesdropper with a predetermined threshold. Subject to the outage probability constraint of the legitimate user, the optimal 3D deployment of the static UAV-enabled jammer was designed to maximize the jamming coverage. To facilitate a tractable analysis, a large sample set of discrete eavesdropper locations and binary integer variables were introduced to represent the target area and the jamming coverage, respectively. To overcome the non-convexity of the problem, a sub-optimal solution in an alternating manner was proposed by considering the multiple circles placement problem in each iteration. It is worth noting that different definitions of the jamming coverage will lead to different design of transmission schemes.

\section{UAV-Enabled Hybrid Secure Transmission Scheme}

\begin{figure}[t]
\centering
\includegraphics[width=12cm,height=9cm]{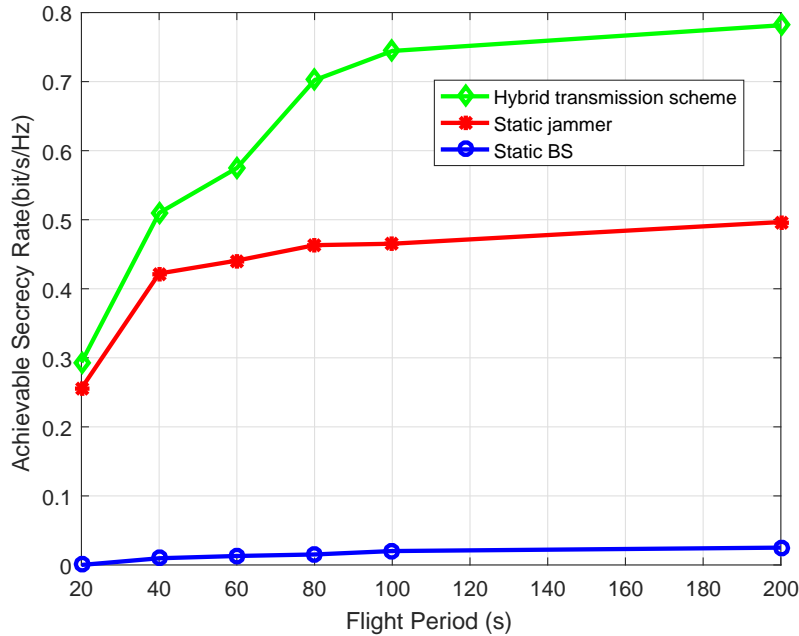}
\caption{The achievable secrecy rate of the scheduled legitimate user against potential internal eavesdropping versus the flight period of UAV aerial base station and/or UAV-enabled friendly jammer. The maximum speed of UAV is 10m/s, the reference SNR at the reference distance ${{d}_0} = 1m$ is 80dB, and the flying altitude of UAV is set as 50m. The locations of $K = 4$ users are ${\mathop{\rm Pos}\nolimits} \;{\rm{ = }}\;\left\{ {\left( {40,30} \right),\left( { - 30,20} \right),\left( { - 20,40} \right),\left( {10, - 30} \right)} \right\}$. The location of static friendly jammer or base station is set as (0,0).}\label{P12}
\end{figure}

Under the scenario involving multiple UAVs for secure communications, different UAVs can be assigned different tasks to provide a hybrid transmission scheme. In \cite{UAVJammingPLS6}, UAV-enabled friendly jamming was exploited to assist secure communications between the UAV aerial base station and multiple legitimate ground users. The time division multiple access (TDMA) protocol was adopted that one legitimate user was scheduled in each time slot while other unscheduled users were treated as potential eavesdroppers. In such a scenario, the worst-case average secrecy rate among legitimate users was maximized and the optimization problem was solved in an alternating iterative manner. In each iteration, the closed-form expressions of the user scheduling binary integer variables were obtained given fixed transmit powers and trajectories of dual UAVs, and then SCA methods were exploited to deal with the non-convexity of other sub-problems. In \cite{UAVJammingPLS7}, a similar idea was investigated in the presence of multiple external ground eavesdroppers. Given the constant transmit power, the collision avoidance constraint was further taken into consideration for the safety of dual UAVs. To deal with the non-convexity caused by the user scheduling binary integer variables, the discrete binary user scheduling constraints were firstly transformed into the equivalent equality constraints by introducing auxiliary continuous variables. The penalty concave-convex procedure (P-CCCP) method was then exploited to solve the obtained problem by incorporating the corresponding penalty terms into the objective function in a double-loop manner. It has been indicated that this design could be directly extended to the general scenarios where there exist multiple UAV friendly jammers.

Compared with other transmission schemes with static friendly jammer or base station, fig. \ref{P12} validates the significant secrecy performance improvement brought by hybrid transmission scheme, due to the mobility characteristics of UAVs. So far, the corresponding researches related to UAV-enabled hybrid secure transmission are limited, which is still an open issue.


\section{Research Challenges and Future Directions}
In this section, some open issues of research challenges and future directions on UAV-involved secure wireless communications are highlighted.

To reduce the complexity of trajectory design, the existing works on this subject roughly fall into two categories: static 3D deployment design under mixed LoS/NLoS conditions and horizontal trajectory design with minimum allowable flight height to avoid collision under dominant LoS scenarios. However, static UAV deployment design is an asymptotic case of dynamic trajectory design with unlimited velocity given predetermined initial and final locations (e.g. for energy charge), and is unable to efficiently adapt to the dynamic environment. On the other hand, dominant LoS scenarios is oversimplified and thus impractical, especially in the suburban, urban and metropolitan areas. Therefore, probabilistic LoS/NLoS channels should be accurately modeled to reflect the practical characteristics of different actual communication environments. Moreover, the distribution of obstacles should be taken into consideration, leading to the practical constraints to design optimal UAV 3D trajectory. In such a context, effective approaches are urgently needed to overcome the high complexity introduced by practical channel models and constraints as aforementioned.

Due to their cost-effective and payload-limited characteristics, multiple UAVs are promising to collaboratively carry out complicated tasks in the future, which is expected to further enhance secrecy performance of UAV-involved secure communications. However, there exist significant variations among different secure communication scenarios, such as the wireless transmission environment, the distribution of legitimate terminals and potential eavesdroppers/attackers, and the security performance objective of concern, Therefore, the optimal configuration of multiple UAVs can be totally different. In such a context, the jointly distributed processing of multi-UAV secure communications combined with optimal configuration design of UAVs for each specific scenario deserves further investigations. On the other hand, due to the scarcity of public spectrum, some currently existing techniques, such as cognitive radio (CR) technology and non-orthogonal multiple access (NOMA) transmission scheme, are supposed to be introduced for spectral efficiency enhancement of multi-UAV secure communications, and the resulted unique security issues should also be taken into consideration.

With the rapid development of UAV manufacturing and antenna miniaturization, UAVs can be equipped with a certain number of antennas in spite of their limited payload capacity. The introduced spatial diversity gain can be exploited for beamforming design, so as to further improve secrecy performance, especially for the low-altitude UAVs with probabilistic multi-path NLoS transmissions. In addition, it is worth noting that the newly-introduced spatial DoFs by multi-antenna technology effectively compensate for the location-dependent DoFs of UAVs, and thus the trajectory design can be simplified under the scenarios where there are multiple related practical constraints. Nevertheless, the performance of multi-antenna transmission depends on preliminary accurate channel estimation results, which are rather difficult to obtain due to high velocity of UAVs and need additional investigations.

Due to limited power storage capability of UAVs, their related energy issues are particularly critical. In such a context, one of the significant goals of UAV-involved secure communications is to simultaneously improve secrecy and energy performances. Since the energy issues associated with UAVs depend on their movement features, such as velocity and acceleration, the resulted trajectory designs are quite complicated, and thus effective approaches are needed to deal with the proposed problems. On the other hand, energy harvesting technology can also be exploited to increase the endurance of UAVs by providing a wireless power supplement. Moreover, it is worth noting that the transmitted energy-bearing signals are information-independent, which can be equivalently designed as jamming signals to simultaneously degrade the quality of wiretapping channels, and thus secrecy performance can be further improved.

Compared with traditional confidential transmissions, which prevent the target messages from being successfully decoded by potential eavesdroppers, covert communications can essentially improve secrecy performance with higher requirements. For UAV-involved covert communication scenarios, the major requirement is to prevent the target messages from interception since it is difficult to hide UAVs. Similar to traditional confidential transmissions, it is expected that UAVs can play multiple roles in covert communications as aforementioned, however, the corresponding research is still limited so far. Therefore, it is a potential research direction to design the configuration of UAVs and optimize the transmit power and/or UAV trajectory for covert communications.

\section{Conclusion}
In this article, we have provided an overview of recent research efforts on UAV-involved secure wireless communications. The typical application scenarios have been categorized as UAV-enabled secure communications and UAV-aided secure cooperation according to different roles of UAVs. Then, the main design considerations including 3D position design of static UAVs, trajectory design of mobile UAVs, and transmit power design have been investigated for secrecy performance enhancement. Finally, we have also provided some valuable research directions on this topic. It is hoped that this overview will lead to more significant and practical researches for UAV-involved secure wireless communications in the future.

\begin{IEEEbiography}[{\includegraphics[width=1in,height=1.25in,clip,keepaspectratio]{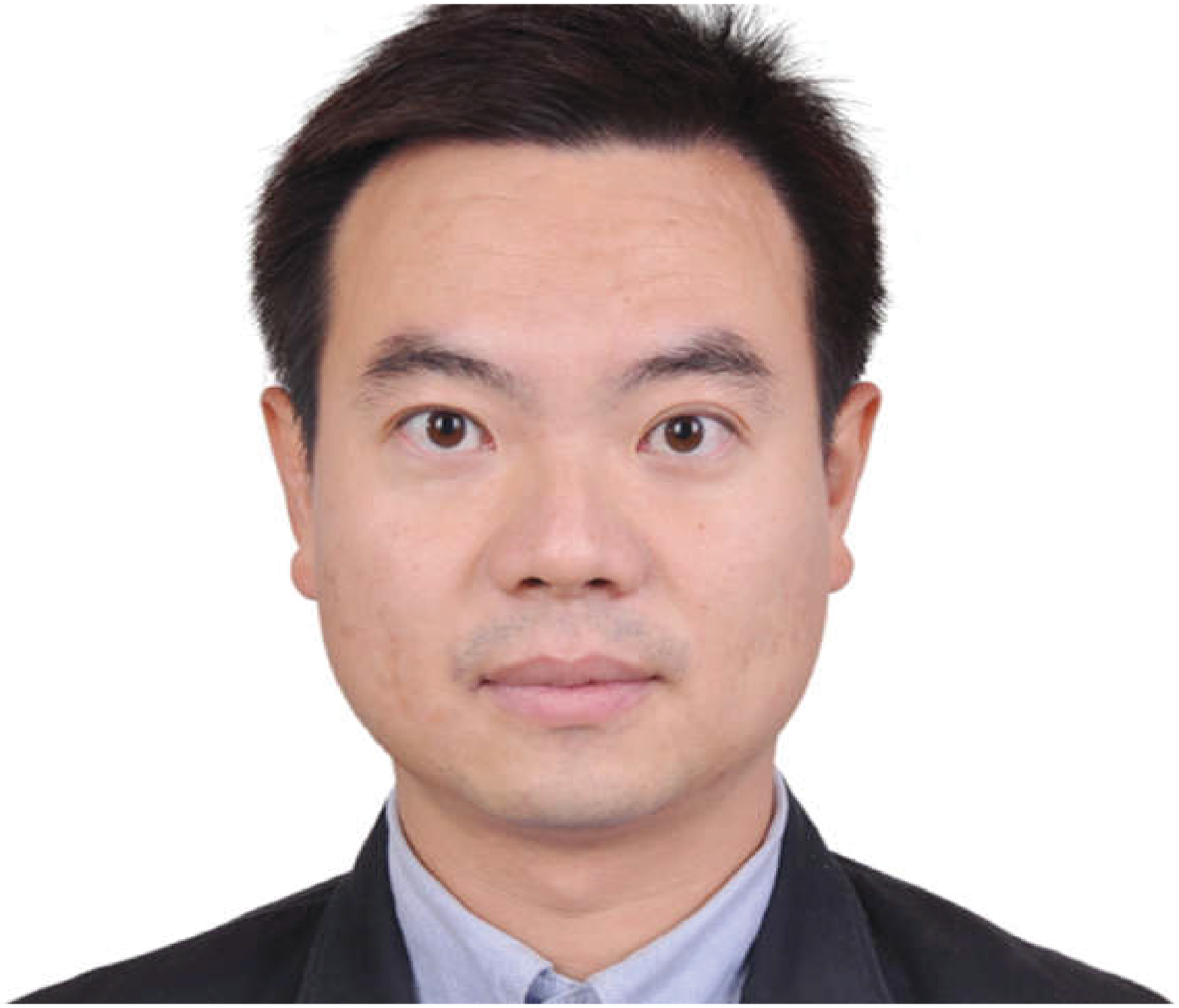}}]
{Hui-Ming Wang}
(S'07--M'10--SM'16) received the B.S. and Ph.D. degrees in electrical engineering from Xi'an Jiaotong University, Xi'an, China, in 2004 and 2010, respectively. From 2007 to 2008, and 2009 to 2010, he was a Visiting Scholar at the Department of Electrical and Computer Engineering, University of Delaware, USA. He is currently a Full Professor with Xi'an Jiaotong University, China. His research interests include 5G communications and networks, physical-layer security of wireless communications, and covert communications. He has co-authored the book \emph{Physical Layer Security in Random Cellular Networks} (Springer, 2016), and has authored or co-authored over 120 IEEE journal and conference papers. He received the IEEE ComSoc Asia-Pacific Best Young Researcher Award in 2018, the National Excellent Doctoral Dissertation Award in China in 2012, and the Best Paper Award from the IEEE/CIC International Conference on Communications in China in 2014. He is currently an Associate Editor of the \emph{IEEE Transactions on Communications}.
\end{IEEEbiography}

\begin{IEEEbiography}[{\includegraphics[width=1in,height=1.25in,clip,keepaspectratio]{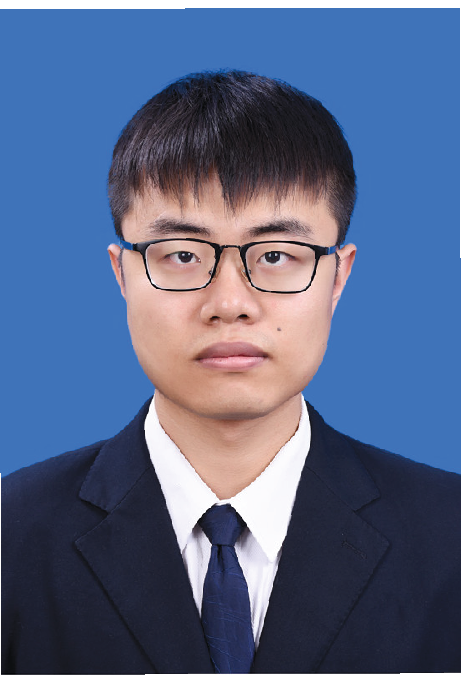}}]
{Xu Zhang} received the B.S. degree in information engineering from Xi'an Jiaotong University in 2017, Xi'an, China. He is currently pursuing the M.S. degree with the Department of Information and Communications Engineering, Xi'an Jiaotong University, and also with the Ministry of Education Key Laboratory for Intelligent Networks and Network Security, China. His current research interests include 5G networks, physical-layer security of wireless communications, and convex optimization.
\end{IEEEbiography}

\begin{IEEEbiography}[{\includegraphics[width=1.05in,height=1.3in,clip,keepaspectratio]{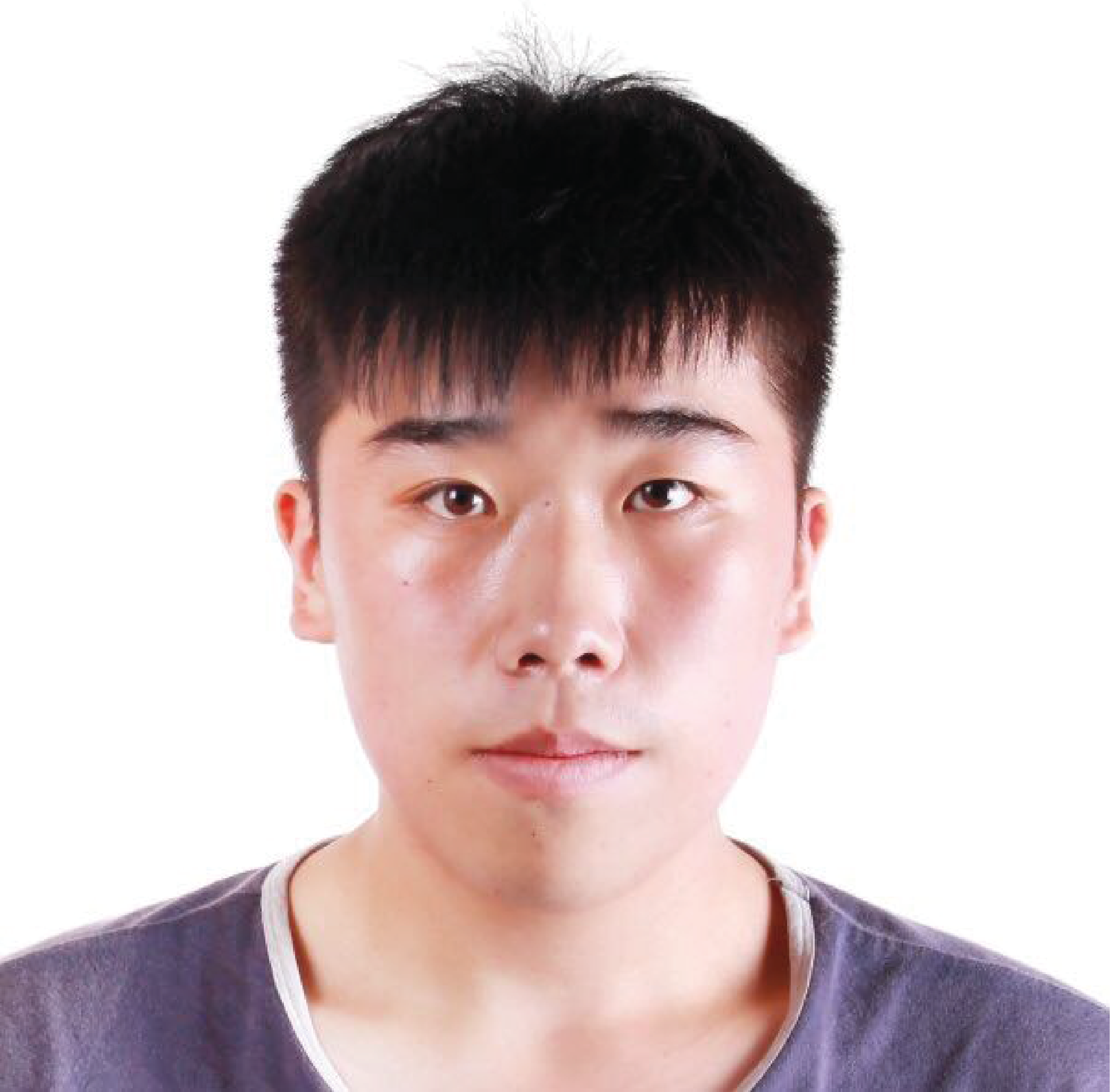}}]
{Jiacheng Jiang} received the B.S. degree in information engineering from Xi'an Jiaotong University, Xi'an, China, in 2018. He is currently working toward the Ph.D degree with the Department of Information and Communications Engineering, Xi'an Jiaotong University, and also with the Ministry of Education Key Laboratory for Intelligent Networks and Network Security, China. His research interests include physical-layer security techniques, wireless channel acquisition, and multiple-input multiple output (MIMO) techniques.
\end{IEEEbiography}

\end{document}